
\magnification=1200
\hsize 15true cm \hoffset=0.5true cm
\vsize 23true cm
\baselineskip=15pt

\font\grande=cmr10 scaled \magstep4
\font\medio=cmr10 scaled \magstep2
\outer\def\beginsection#1\par{\medbreak\bigskip
      \message{#1}\leftline{\bf#1}\nobreak\medskip\vskip-\parskip
      \noindent}

\def \me {\buildrel <\over \sim}
\def \Me {\buildrel >\over \sim}

\def \pr {\prime}

\def \b {\beta}
\def \a {\alpha}

\def \ga {\gamma}
\def \sg {\sigma}
\def \Sg {\Sigma}

\def \om {\omega}
\def \Om {\Omega}
\def \noi {\noindent}

\def\sqr#1#2{{\vcenter{\hrule height.#2pt\hbox{\vrule width.#2pt
height#1pt \kern#1pt\vrule width.#2pt}\hrule height.#2pt}}}

\def\lsim{\mathrel{\rlap{\lower4pt\hbox{\hskip1pt$\sim$}}
    \raise1pt\hbox{$<$}}}         
\def\gsim{\mathrel{\rlap{\lower4pt\hbox{\hskip1pt$\sim$}}
    \raise1pt\hbox{$>$}}}         

\nopagenumbers
\line{\hfil CERN-TH.6836/93}
\line{\hfil DFTT-26/93}
\line{\hfil March 1993}

\vskip 1.5 true cm

\centerline{\grande Squeezed Thermal Vacuum}
\bigskip
\centerline{\grande and the Maximum Scale for Inflation}

\vskip 1true cm
\centerline{M. Gasperini \footnote{*}
{Permanent address:  Dipartimento di
Fisica Teorica, Via P.Giuria 1, 10125 Turin, Italy}}
{\it
\centerline {Theory Division, CERN, Geneva, Switzerland}}

\centerline{M. Giovannini}
{\it
\centerline {Dipartimento di Fisica Teorica, Via P.Giuria 1,
10125 Turin, Italy}
\centerline {and INFN, Sezione di Torino, Turin, Italy}}
\centerline {and}
\centerline {G. Veneziano}
{\it
\centerline {Theory Division, CERN, Geneva, Switzerland}}

\vskip 1 true cm

\baselineskip=15pt
\centerline{\medio Abstract}
\noi
We consider the stimulated emission of gravitons from an initial state of
thermal equilibrium, under the action of the cosmic gravitational
background field.
We find that the low-energy graviton spectrum is enhanced if
compared with spontaneous creation from the vacuum; as a
consequence, the scale of inflation must be lowered, in order not to exceed
the observed CMB quadrupole anisotropy. This effect is particularly important
for models based on a symmetry-breaking transition which
require, as initial condition, a state  of thermal  equilibrium at
temperatures of the order of the inflation scale.
\vskip 1 true cm
\centerline{ To appear in {\bf Phys. Rev. D} {\it (Rapid
Communications)}}

\vfill\eject

\footline={\hss\rm\folio\hss}

\pageno=1

\baselineskip=20pt
It has recently been argued that the Universe cannot be in a state
of ``eternal'' inflation: not only a primordial de Sitter exponential
expansion, without a beginning in time, is impossible [1], but also
there are  quantum cosmological arguments [2] suggesting for
the inflationary phase just the minimum duration required to
bring in causal contact scales that are not much larger than
our present Hubble radius $ H_{0}^{-1}$.

If this is the case, the initial state, whose perturbations are amplified by
the
subsequent inflationary evolution, should be
fixed by the
dynamics of the pre-inflationary era and could
differ considerably from the
usually assumed vacuum.
This difference influences the shape
of the perturbation spectrum, as first pointed out in [3] for the case
of tensor perturbations (graviton production), and recently
discussed in [4] for the
scalar perturbation case.

It is well known, in particular, that any inflationary model based on a
temperature dependent phase transition necessarily requires a homogeneous
thermal state as initial condition. The existence of initial thermal
equilibrium constrains the parameters of such models, and a recent
discussion [5] suggests that, in their context, a sufficient duration
of inflation can be arranged only for low enough energy scales. The aim
of this paper is to point out that also
the conventional upper bounds on the scale
arising from the gravity-wave contribution to the CMB quadrupole
anisotropy are to be lowered, if
the relic graviton background is produced
from an initial thermal bath rather than from the zero-temperature
vacuum.

Indeed, the inflationary amplification of the vacuum fluctuations can be
properly represented, in second quantized formalism, as a process of pair
production under the action of the external gravitational field [6], with
the produced particles necessarily appearing in a final ``squeezed
vacuum''
quantum state [6,7]. If the initial vacuum is changed into a state of
thermal equilibrium, while the inflationary dynamics is left unchanged,
the pair production process leads eventually to a thermal mixture
of``squeezed number'' states, instead of the pure squeezed vacuum. This
modifies the spectral number density of the produced particles, in
such a way that their energy distribution deviates in general from
the flat Harrison-Zeldovich spectrum, even in the case of pure
de Sitter inflationary dynamics. The isotropy properties of the CMB
radiation then provide a bound on the inflation scale, which
depends on the temperature of the initial thermal bath.

In order to discuss this effect let us recall [6--8], first of all,
that the inflationary particle production can be described in terms
of Bogoliubov transformations relating, for each mode $k$,
the $|\rm{in}\rangle$ ($b_{k},b_{k}^{\dagger}$)
to the $|\rm{out}\rangle$
($a_{k},a_{k}^{\dagger}$)
annihilation and creation operators :
$$
a_{k}=c_{+}(k)b_{k}+c_{-}^{\ast}(k)b_{k}^{\dagger}~~~,~~~
a_{k}^{\dagger}=c_{-}(k)b_{k}+ c_{+}^{\ast}(k)b_{k}^{\dagger}. \eqno(1)
$$
The Bogoliubov coefficients $c_{\pm}(k)$ depend on the dynamics of the
background geometry (in particular on  the transition from the inflation
to the standard decelerated phase), and satisfy $\mid c_{+}\mid ^{2} -
\mid c_{-} \mid^{2}~=1$. By parametrizing $c_{\pm}$ as:
$$
c_{+}(k)=\cosh r_{k}~~~~~,~~~~~c_{-}^{\ast}=e^{2i\vartheta_{k}}\sinh
r_{k},
\eqno(2)
$$
the relations (1) can be rewritten as unitary transformations
$$
a_{k}=\Sg_{k} b_{k}\Sg_{k}^{\dagger}~~~~,~~~~a_{k}^{\dagger}=
\Sg_{k}b_{k}^{\dagger}\Sg_{k}^{\dagger} \eqno(3)
$$
generated by the squeezing operator \footnote{*}{
Note that, although the correct (momentum conserving) transformation
should involve the two-mode squeezing operator [9], we are considering
here the simpler one-mode formalism since it gives completely
equivalent results for the quantities discussed in this paper.}:

$$
\Sg_{k}=\exp[(z_{k}^{\ast}b_{k}^{2}-z_{k}{b_{k}^{\dagger}}^{2})/2]~~~~,~~~~
z_{k}=r_{k}e^{2i\vartheta_{k}}. \eqno(4)
$$

The spectral properties of the relic radiation are usually
derived by starting from the $ |\rm{in}\rangle$ vacuum state $ |0
\rangle$, which
satisfies
$b_{k}|0\rangle=0$.
The pair production then process leads, for each mode, to the squeezed
vacuum state $|z_{k}\rangle =\Sg_{k}|0\rangle$, such that
$a_{k} |z_{k}\rangle =0$. The average particle number can be expressed in
terms of
the squeezing parameter $r_{k}$ as
$$
\overline N_{k}=~\langle 0| a^{\dagger}_{k}a_{k}| 0\rangle~=~| c_{-}(k)|
^{2}= \sinh^{2}r_{k}.   \eqno(5)
$$
If we start, however, with a number state $| n_{k}\rangle$ in which $n$
particles are already present in the given mode,
$b_{k}^{\dagger}b_k|n_{k}\rangle=n_{k}|n_{k}\rangle$, we obtain a
``squeezed number'' [10] state $| z_{k},n_{k}\rangle=\Sg_{k}| n_{k}
\rangle$,
with
$$
\overline N_{k}=\langle n_{k}| a_{k}^{\dagger}a_{k}| n_{k}\rangle~=~
|c_{-}|^{2} (1+n_{k})+n_{k}(1+| c_{-}|^2), \eqno(6)
$$
in agreement with the rules of stimulated emission.
If we start, more generally, with a statistical mixture,
$$
\rho_{k}=\sum_{n}  P_{n}(k) | n_{k}\rangle \langle n_{k}|~~~~,
{}~~~~\sum_{n}P_{n}=1 ,
\eqno(7)
$$
then [11]
$$
\overline N_{k}=\rm{Tr}(\rho_{k}a_{k}^{\dagger}a_{k})=
|c_{-}(k)|^2(1+ \overline n_{k})+
\overline n_{k}(1+ |c_{-}(k)|^{2}), \eqno(8)
$$
where $\overline n_{k}=\sum_{n} n P_{n}(k)$ is the initial averaged particle
number. The spectral number distribution for the (bosonic) massless
particles produced from an initial thermal bath, at a co-moving
temperature $\overline T=\overline\beta^{-1}$, is thus obtained by
inserting into eq.(8) the thermal average number
$$
\overline n_{k}=(e^{\overline\beta k}-1)^{-1}.  \eqno(9)
$$

This result can be checked directly by explicit use of the ``squeezed
thermal'' density operator [10], $\rho_{st}=\Sg\rho_{t}\Sg^{\dagger}$,
where $\rho_{t}=\exp(-\overline\beta b^{\dagger}b)$ (from now on the mode
index is to be understood, if not explicitly written; any correlation
among modes is moreover neglected, as we are treating perturbations
in the linear approximation).
Indeed, in the convenient representation defined by the ``superfluctuant''
variable $ x$,
whose variance $\Delta x$ is amplified by the squeezing process [9],
$\rho_{st}$ takes the form:
$$
\rho_{st}(x,x^{\prime})=~\langle x|\rho_{st}| x^{\pr}\rangle~=
$$
$$
=[{{\pi(2\overline n+1)}\over
\sg}]^{-1/ 2} \exp\{-{{2\overline n(\overline n +1) +1}\over
{2(2\overline n +1)}}(x^{2}+{x^{\prime 2}})\sg+
{{2\overline n(\overline n +1)}\over {(2\overline
n+1)}}xx^{\prime}\sg\},
 \eqno (10)
$$
where $\sg=\exp(-2r)$, and $\overline n$ is the thermal average
number of eq. (9) (for $\sg=1$, eq. (10) reduces to the usual
thermal density matrix in the configuration space representation).
The computation of $\rm{Tr}(\rho_{st}b^{\dagger}b)$ then reproduces exactly
eq. (8), with $| c_{-}|^{2}~=\sinh^{2}r$.

Note, incidentally, that, unlike $\overline N$,
the entropy growth $\Delta S$ associated with pair production is not
affected by finite-temperature corrections to the
initial state; for a squeezed
thermal mixture, $\Delta S$ turns out to be just the
same as that obtained starting from the vacuum [12,13], namely
$\Delta S=-\ln \sg=2r$.
This can be easily verified by computing $-\rm{Tr} \rho_{st}\ln \rho_{st}$
for the two-mode generalization of the  matrix (10), and by
subtracting the initial thermal contribution.

It is also worth noting that, in the real-time formalism
of the  thermo-field dynamics [14], the thermal vacuum is related to the
$T=0$ vacuum by a kind of
Bogoliubov transformation, with squeezing parameter
$r_{t}=\sinh^{-1} [(e^{\overline\beta k} -1)^{-1/ 2}] $.
Such a transformation acts on a ``doubled'' Hilbert space, obtained by
introducing fictitious operators associated with each physical operator.
In this context, the
average number of eq. (8) can be recovered, formally,
by considering the state obtained from
the vacuum by the product of two $SU(1,1)$
Bogoliubov matrices, with
parameters $r_{1}=r_{t}$, and $r_{2}=\sinh^{-1} | c_{-}|$, provided
the relative phase is chosen to be
$\vartheta_{1}-\vartheta_{2}={{(2m+1)\pi}/ 4} $,
with $m$ integer.

We shall now concentrate, in particular, on the stimulated emission of
gravitons from an initial thermal bath, under the action of a changing
background geometry which describes the transition from an
inflationary phase to the subsequent radiation-dominated
and matter-dominated eras.

The spectral energy density $\rho(\om)$, which is the variable
usually adopted to characterize today's  distribution of the
produced gravitons [8,15,16] is given by
$$
\rho(\om)=\om({{d\rho_{g}}\over d\om})\simeq\om^{4} \overline N(\om),
\eqno(11)
$$
where $\overline N$ is defined in eq. (8), with the corresponding
$\overline n$ of eq. (9) (we have neglected numerical factors of
order unity, and $\om$ is the proper frequency, related to the
comoving one, $k$, by $\om ={k/ a(t)}$, where $a$
is the scale factor of the background isotropic metric).

The Bogoliubov coefficients $c_{\pm}(\om)$, connecting the $|\rm{
in}\rangle$
and  $| \rm{out}\rangle$
graviton modes for the inflation $\rightarrow$ radiation $\rightarrow$
matter transition, have been computed by many
authors [8,16,17], in the sudden approximation.
In such an approximation, one ignores the details
of the transition among the three different cosmic phases, and
the particle production is neglected for modes which never ``hit''
the effective potential barrier appearing in the graviton wave equation.
As a consequence, the Bogoliubov coefficients are not modified, in this
approximation, if an initial phase dominated by a thermal radiation bath
is inserted before the de Sitter era, since the radiation
dominated evolution gives no contribution to that potential barrier.

We then insert the known expression of $c_{-}(\om)$ in eq. (11)
and measure $\rho(\om)$, as usual, in units of critical energy density
$\rho_{c}$, defining $\Om(\om)={{\rho(\om)}/\rho_{c}}$.
By exploiting the fact that $|c_{-}|\geq 1$ for all the
modes undergoing the parametric amplification, we finally get
(we follow in particular the notations of [17]):
$$
\Om(\om,t_{0})\simeq G{H_{1}}^{2} \Om_{\ga}(t_{0})({\om\over\om_{1}})^
{2-2\a} \coth({\b_{0}\om\over 2})~~~~~,~~~~~~\om_{2}<\om<\om_{1}
$$
$$
\Om(\om,t_{0})
\simeq G {H_{1}}^{2}\Om_{\ga}(t_{0})({\om\over\om_{1}})^{2-2\a}
({\om\over\om_{2}})^{-2}
\coth({\b_{0}\om\over 2})~~~,~~~~~\om_{0}<\om<\om_{2}.  \eqno(12)
$$
Here ${\b_{0}}^{-1}\equiv \b^{-1}(t_{0})$ is the proper
temperature of the initial thermal state, adiabatically rescaled
down to the present observation time $t_{0}$ [$\b_{0}$ is defined
in terms of the comoving temperature $\overline\b$ as
$\b(t_{0})=\overline\b a(t_{0})$]; $\Om_{\ga}(t_{0})\simeq 10^{-4}$
is the fraction of the critical energy density present today in the
form of radiation; $\a\ge 1$ is a coefficient parametrizing
(in conformal time) the power-law behaviour of the scale factor;
$H_{1}\equiv H(t_{1}) $ is the curvature scale at the time $t_{1}$
marking  the end of inflation  and the
beginning of the radiation-dominated era; $\om_{0}\simeq 10^{-18}$ Hz is
the minimum amplified frequency crossing today the Hubble
radius $H_{0}^{-1}$; $\om_{2}\simeq 10^{2} \om_{0}$ is the
frequency corresponding to the matter radiation transition; $\om_{1}$,
finally, is the maximum amplified frequency,
related to the inflation scale by $\om_{1}\simeq 10^{11} ({H_{1}/
M_{P}})^{1/ 2} $ Hz ($M_{P}$ is the Planck mass).

Equation (12) provides
the present energy distribution of a gravity-wave
background of inflationary origin, obtained from a
primordial state of
thermal equilibrium at a proper temperature $\b^{-1}$. In the limit
$\b\rightarrow\infty$, and for $\a=1$, we recover the well-known flat
de Sitter spectrum obtained from the vacuum, with the usual frequency
dependence ($\sim\om^{-2}$) at low energy, due to the radiation-matter
transition [15--17].

The primary effect of the initial finite temperature is to
enhance the low-frequency graviton production, with respect to
the high-frequency sector of the spectrum.
In this respect, the initial thermal vacuum mimics the effect
of putting ``more power on larger scales'' [18], typical of
power-law inflation; with the difference, however, that the
thermal effects are rapidly damped at high $\om$, for realistic values
of the rescaled initial temperature $\b_{0}$.
In  Fig.1 the spectrum has been plotted for the de Sitter case
($\a=1$), in such a way as to represent, for various values of $\b_{0}$,
the maximum allowed fraction of critical energy density
compatible with the CMB isotropy.

The relic graviton spectrum is mainly constrained, indeed, by three
kinds of direct observations [15]: CMB isotropy, pulsar timing data,
and critical density. However, as discussed in [15] and [17],
the most significant constraint for flat or decreasing spectra, such as
those of eq. (12), turns out to be the isotropy
bound imposed at the minimum frequency $\om_{0}$, where it presently
implies
$\Om(\om_{0},t_{0})\me 10^{-10}$ (we are making here a conservative use of
the COBE data [19], as an upper limit on the graviton contributions to
the quadrupole anisotropy). Such a condition, imposed on eq. (12),
provides a bound on the inflation scale $H_{1}$, which can be
conveniently expressed, in terms of the usual spectral index
$n=3-2\a$,
$$
\rm{log}_{10}({H_{1}\over M_{P}})\me {2\over {3+n}}\left(29n-39+
\rm{log}_{10} (\tanh{{\b_{0}
\om_{0}}\over 2})\right). \eqno(13)
$$

In the limit $\b_{0}\rightarrow\infty$ this generalizes, to any value of
$n$, the usual isotropy constraint on the curvature scale of de Sitter
($n=1$) inflation [20], namely $H_{1}\me 10^{-5} M_{P}$ (with an
improvement of one order of magnitude with respect to [15--17,20],
due to the use of the more constraining COBE data). The new effect,
however, is that for finite initial temperature the maximum allowed
scale is in general depressed with respect to vacuum production, as
illustrated in Fig.2 (for $\b_{0}\om_{0}\ll 1$, in particular,
$H_{1}$ scales like $\b_{0}^{2/{(3+n)}}$). The inclusion of the
thermal correction is thus expected to modify
the existing relations (see e.g. [21])
between the power index and the scale of inflation, obained
by fitting the observed anisotropy on a $10$
degree
angular scale.

In particular,
for the inflationary models based on a thermal symmetry breaking
mechanism, the initial temperature $\b^{-1}$ is not
independent from the scale
of inflation itself. Suppose, indeed, that the inflationary
phase transition occurs at an energy scale $M$, which is the scale at
which the time-independent vacuum energy becomes dominant ($M$ is related
to the curvature scale $H_{1}$ by ${M/ M_{P}}=\left({H_{1}/
M_{P}}\right)^{1/ 2}$, according to the Einstein equations). At
earlier times,  such that $\b^{-1}(t)>M$, the symmetry is restored, and the
Universe becomes radiation-dominated. The temperature of the initial
thermal ensemble, rescaled at the beginning ($t=t_{i}$) of inflation,
must therefore satisfy $\b(t_{i})M\leq 1$.

This condition, rescaled down adiabatically at the present time $t_{0}$,
provides a bound on the spectral parameter $\b_{0}$, which depends on the
duration $Z$ of the inflationary phase ($Z={a(t_{1})/ a(t_{i})}$,
where $t_1$ marks the end of inflation):
$$
\b_{0}\om_{0}=M\b_{i}\left(a_{0}\over a_{i}\right)\left(\om_{0}\over
M\right)\leq Z\left(H_{0}\over M_{P}\right)^{1\over 2}.  \eqno(14)
$$

For $Z\rightarrow\infty$ this bound is washed out by the inflationary
supercooling of the original thermal ensemble. However, for models
whose parameters are adjusted  to give just the minimal amount of
inflation required to solve the standard problems [22],
$$
Z\simeq Z_{min}= e^{53}\left(M\over {10^{14}\rm{GeV}}\right)^{2\over 3}\left(
T_{rh}\over {10^{10}\rm{GeV}}\right)^{1\over 3}  \eqno(15)
$$
($T_{rh}$ is the reheating temperature), the bound (14) can be
re-expressed in terms of the reheating efficiency, $Q={T_{rh}/ M}$,
as
$$
2~\rm{log}_{10}(\b_{0}\om_{0}) \me \rm{log}_{10}
\left(H_{1}\over M_{P}\right)+{2\over 3}\rm{log}_{10}Q.
\eqno(16)
$$
For each given value of $n$ and $Q$ one has then a minimum allowed
temperature $\b_{0}^{-1}$ and a maximum allowed scale $H_{1}$, which are
fixed by the combination of the constraints (13) and (16), as
illustrated in Fig.3 for three different spectral indices. This
effect represents a truly ``remnant'' of
the pre-inflationary Universe, in the sense
of [22]. For $Z>Z_{min}$, the maximum allowed value of $H_1$ scales up
like $Z^{2/(3+n)}$ and becomes $Z$-independent for $\b \om_0 \Me 1$.

In conclusion, we want to stress that the numerical results presented
here should be regarded, in many respects, as semi-quantitative results,
because of the approximations made and of the uncertainty of the
experimental data, which has not been completely  taken into account
in our discussion. Nevertheless, we believe that already at this
qualitative  level two important indications emerge rather clearly.

The first is that {\it any fit} of the CMB anisotropy in
terms of the gravity wave background {\it should include a thermal
dependence in the spectrum} [according to eq. (12)], in order to take
into account the possible finite temperature of the initial state.

The second indication is that, even if a thermal phase transition at the
GUT scale is certainly not ruled out as a
possible inflationary mechanism,
the effects discussed here seem to provide some motivation for
investigating the same mechanism also at lower scales, such as the
electroweak one [23]. The constraints on the scale imposed by our
results
seem to challenge, in
particular,  thermal models of inflation obtained in the context of
string theory, where the natural scale is very near the Planck one. No
difficulty seems to arise, on the contrary, in
the context of a ``duality-symmetric'' string cosmology [24] where the
inflationary phase is the dual counterpart of the
present decelerating expansion, and starts naturally from a flat
and cold low-energy vacuum.

\vfill\eject

\centerline{\bf References}

\item{1.}A.Vilenkin, Phys.Rev.D 46, 2355(1992)

\item{2.}L.P.Grishchuk, Phys.Rev.D., 4717(1992)

\item{3.}L.P.Grishchuk and Y.V.Sidorov, Class.Quant.Grav.6, L155(1989)

\item{4.}I.Y.Sokolov, Class.Quant.Grav.9, L61(1992)

\item{5.}P.D.B.Collins and R.F.Langbein, Phys.Rev.D 45, 3429(1992)

\item{6.}L.P.Grishchuk and Y.V.Sidorov, Phys.Rev.D 42, 3413(1992);

L.P.Grishchuk, ``Quantum mechanics of the primordial cosmological
pertur-

bations'', in Proc. 6th. Marcel Grossmann Meeting (Kyoto, June
1991);

V.F.Mukhanov, H.A.Feldman and R.H.Brandenberger, Phys.Rep. 215, 203

(1992)

\item{7.}L.P.Grishchuk and Y.V.Sidorov, Class.Quant.Grav.6, L161(1989);

L.P.Grishchuk, ``Squeezed states in the theory of primordial gravitational

waves'', in Proc. Workshop on Squeezed States and Uncertainty

Relations (Maryland Univ.), eds. D.Han, Y.S.Kim and W.W.Zachary

(NASA Conf. Pub. No. 3135, 1992) p. 329

\item{8.}L.P.Grishchuk and M.Solokhin, Phys.Rev.D 43,2566 (1991)

\item{9.}B.L.Schumaker, Phys.Rep.135,317(1986)

\item{10.}M.S.Kim. F.A.M.de Oliveira and P.L.Knight,
in ``New frontiers in quantum electrodynamics and quantum optics'', ed.
A.O.Barut (Plenum, New York, 1990) p. 231

\item{11.}L.Parker, Phys.Rev.183, 1057 (1969)

\item{12.}M.Gasperini and M.Giovannini, Phys.Lett.B301, 334 (1993)

\item{13.}R.Brandenberger, V.Mukhanov and
T.Prokopec, Phys.Rev.Lett. 69, 3606

(1992);
The entropy of gravitational field, Brown-Het-849 (August 1992);

T.Prokopec, Entropy of the squeezed vacuum, Brown-Het-861 (June 1992)

\item{14.}H.Matsumoto,M.Tachiki and H.Umezawa, Thermo field dynamics
(North Holland, Amsterdam, 1982)

\item{15.}L.P.Grishchuk, Sov.Phys.Usp.31, 940(1988)

\item{16.}B.Allen, Phys.Rev.D 37,2078(1988);

V.Sahni, Phys.Rev.D 42, 453(1990);

M.Gasperini and M.Giovannini, Phys.Lett.B 282, 36(1992)

\item{17.}M.Gasperini and M.Giovannini, Phys.Rev.D47, 1529 (1993)

\item{18.}M.S.Turner, in Proc. First Erice School ``D.Chalonge'' on
Astrofundamental Physics (September 1991), ed. N.Sanchez (World
Scientific, Singapore).

\item{19.}G.Smoot et al., Astrophys.J.316, L1(1992)

\item{20.}V.A.Rubakov, M.V.Sazhin and A.V.Veryaskin, Phys.Lett.B 115,
189(1982);

R.Fabbri and M.B.Pollock, Phys.Lett.B 125, 445 (1983);

L.F.Abbott and M.B.Wise, Nucl.Phys.B 244,541(1984)

\item{21.}M.White, COBE and the scale of power-law inflation, Berkeley
preprint CfPA-TH-92-033 (November 1992)

\item{22.}M.S.Turner, Phys.Rev.D 44, 3737(1991)

\item{23.}L.Knox and M.S.Turner, Phys.Rev.Lett. 70, 371(1993)

\item{24.}M.Gasperini, N.Sanchez and G.Veneziano,
Nucl.Phys.B364, 365(1991);

G.Veneziano, Phys.Lett.B265, 287(1991);

K.A.Meissner and G.Veneziano, Mod.Phys.Lett.A6, 3397 (1991);

M.Gasperini and G.Veneziano, Phys.Lett.B277,256(1992);

M.Gasperini and G.Veneziano, ``Pre-big-bang in string
cosmology'',

Astroparticle Physics (1993) (in press)

\vfill\eject

\centerline{\bf Figure Captions}

\vskip 2 true cm
\item{{\bf Fig. 1}} Maximum allowed spectral energy density [according to eq.
(12)] in a relic
graviton background, produced after a phase of de Sitter inflation, from
an initial
thermal bath at finite temperature $T=\b^{-1}$. The rescaled temperature
$\b_0^{-1}$ is measured here in units of $\om_0=10^{-18}$ Hz.
\bigskip
\item{{\bf Fig. 2}} Maximum allowed inflation scale versus the spectral
index $n$, according to eq. (13), for three different values of the initial
temperature $\b_0^{-1}$ (in units of $\om_0$).
\bigskip
\item{{\bf Fig. 3}} Maximum scale and minimum initial temperature for models
with the inflation rate of eq. (15), and with reheating efficiency
$Q=1$ and $Q=10^{-3}$. The allowed region lies \underbar {below}
the curves labelled
by the spectral index $n$, and \underbar{to}
\underbar{the left} of the curves labelled by $Q$.

\end